\documentclass[superscriptaddress,pra]{revtex4}

\usepackage[latin1]{inputenc}
\usepackage{graphicx}
\usepackage{amsmath}
\usepackage{amssymb}
\usepackage{dsfont}
\usepackage{graphicx}

\makeatletter
\def\erf{\mathop{\operator@font erf}\nolimits}
\makeatother

\newcommand\be{\begin{equation}}
\newcommand\ee{\end{equation}}

\begin{document}

\title{Optical realization of a quantum beam splitter}
\author{R. Mar-Sarao and H. Moya-Cessa}
\affiliation{INAOE, Apdo. Postal 51 y 216, 72000, Puebla, Pue.,
Mexico}

\begin{abstract}
We show how the quantum process of splitting light may be modelled
in classical optics. A second result is the possibility to
engineer specific forms of a  classical field.
\end{abstract}
\pacs{} \maketitle


\subsection{Introduction}
The modelling of quantum mechanical systems with classical optics
is a topic that has attracted interest recently.  Along these
lines Man'ko \textit{et al.} have proposed  to realize quantum
computation by quantum like systems \cite{Mank} and Crasser
\textit{et al.} \cite{Sch} have pointed out the similarities
between quantum mechanics and Fresnel optics in phase space.
Following these cross-applications, here we would like to show how
a quantum beam splitter may be modelled in classical optics. The
possibility of generating specific forms (engineering) of the
propagated field is also studied.

A beam splitter is an optical component that combines two
propagating modes into two other propagating modes. Fig. 1 shows
the setup that produces this effect for a 50:50 beam splitter. Two
fields $a_x$ and $a_y$ enter the beam splitter and two fields exit
it, that correspond to a combination of the entering ones. In
quantum optics the beam splitter is modelled by the interaction of
two fields \cite{Villas} with the Hamiltonian given by
\begin{equation}
H=\omega_x a_x^{\dagger}a_x+ \omega_y a_y^{\dagger}a_y +
\chi(a_y^{\dagger}a_x+a_x^{\dagger}a_y)
\end{equation}%
where the $\omega$'s are the field frequencies, $\chi$ is the
interaction constant and $a_j$ and $a_j^{\dagger}$ ($j=x,y$) are
the annihilation and creation operators of the field modes.

If we consider equal field frequencies, we can obtain the beam
splitter operator \cite{Campos}
\begin{equation}
B=e^{-i\theta(a_y^{\dagger}a_x+a_x^{\dagger}a_y)}
\end{equation}%
with $\theta=\chi t$. Note that
\begin{eqnarray}
\nonumber &&Ba_xB^{\dagger}=\cos(\theta)a_x+i\sin(\theta)a_y, \\
&& Ba_yB^{\dagger}=\cos(\theta)a_y+i\sin(\theta)a_x
\end{eqnarray}%
such that $\cos(\theta)$ and $\sin(\theta)$ may be related
directly with the transmission and reflection coefficients. One
feature present in the quantum beam splitter is that,  if there is
only one field entering by one of the arms of the beam splitter,
one has to consider always a vacuum field entering by the other
arm. It is well known that this system produces entanglement
\cite{Kim,Scheel}, for instance, if we consider the 50:50 beam
splitter, i.e. we set $\theta=\pi/4$, (see Fig. 1) and in each of
the arms the first excited number state, namely the state
$|\psi_I\rangle=|1\rangle_x|1\rangle_y$ as initial state, we have
as final state
\begin{equation}
|\psi_F\rangle=\frac{i}{\sqrt{2}}(|2\rangle_x|0\rangle_y+|0\rangle_x|2\rangle_y),\label{2photons}
\end{equation}%
this is an entangled state that tells  that both photons travel
together.

\subsection{Modelling field-field interaction}

We consider the paraxial propagation of a field, that has the
equation form (see for instance \cite{Chavez}
\begin{equation}
2ik_{0}\frac{\partial E}{\partial z}=\nabla _{\perp
}^{2}E+k^{2}(x,y)E \label{waveq}
\end{equation}%
where $k_{0}=2\pi n_{0}/\lambda $ is the wavenumber with $\lambda
$ the wavelength of the propagating lightbeam, $n_{0}$ is the
homogeneous refractive index. The function $k^{2}(x,y)$ describes
the inhomogeneity of a medium responsible for the waveguiding of
an optical field $E$. It has been recently shown that by using an
astigmatic and slightly tilted probe beam, it produces a GRIN like
medium, with an extra cross term $\chi xy$ \cite{Chavez}. Taking
into account such inhomogeneity the paraxial wave equation is
written as
\begin{eqnarray} \nonumber
&&i\frac{\partial E}{\partial z}=\frac{\nabla _{\perp
}^{2}}{2k_{0}}E\\&+&\left(
k_{0}/2-\frac{1}{2}(k_{0}\tilde{\alpha}_{x}^{2}x^{2}+k_{0}\tilde{\alpha}%
_{y}^{2}y^{2})+\chi xy\right) E, \label{psi}
\end{eqnarray}%
with $k_{0}\tilde{\alpha}_{q}^{2}, \qquad q=x,y$ inhomogeneity
parameters related to the generated GRIN medium \cite{Chavez}.

\subsection{Optical realization of
the quantum beam splitter} Let us define the \textit{ladder}
operators \cite{Arfken}
\begin{eqnarray}\nonumber
a_{q}=\sqrt{\frac{k_{0}\tilde{\alpha}_{q}}{2}}q+\frac{1}{\sqrt{2k_{0}\tilde{%
\alpha}_{q}}}\frac{d}{dq},\\ a_{q}^{\dagger }=\sqrt{\frac{k_{0}\tilde{%
\alpha}_{q}}{2}}q-\frac{1}{\sqrt{2k_{0}\tilde{\alpha}_{q}}}\frac{d}{dq},
\label{ladder}
\end{eqnarray}%
with $q=x,y$.  These operators are also called  creation and
annihilation operators in quantum optics because the action of
$a_{x}$ on a function
\begin{equation}
u_{m}(x)=\left( \frac{k_{0}\tilde{\alpha}_{x}}{\pi }\right) ^{1/4}\frac{1}{%
\sqrt{2^{m}m!}}H_{m}(\sqrt{k_{0}\tilde{\alpha}_{x}}x)e^{-k_{0}\tilde{\alpha}%
_{x}x^{2}/2} \label{descomp}
\end{equation}%
where $H_{m}(x)$ are Hermite polynomials, gives
\begin{equation}
a_{x}u_{m}(x)=\sqrt{m}u_{m-1}(x)
\end{equation}%
and
\begin{equation}
a_{x}^{\dagger }u_{m}(x)=\sqrt{m+1}u_{m+1}(x).
\end{equation}%
The above equations are also valid for the $y$ coordinate simply
making the change $y\rightarrow x$. Furthermore, note that
$[a_{x},a_{x}^{\dagger }]=1$. We can rewrite (\ref{psi}) using
(\ref{ladder}) as
\begin{eqnarray}
i\frac{\partial E}{\partial z}=\left( \tilde{\alpha}_{x}(n_{x}+\frac{1}{2})+%
\tilde{\alpha}_{y}(n_{y}+\frac{1}{2})+\chi xy+\frac{k_0}{2}\right)
E , \label{two-fields}
\end{eqnarray}%
with $n_{j}=a_{j}^{\dagger }a_{j}$ for $j=x,y$ the so-called number operator. We transform (\ref%
{two-fields}) to get rid of the constant terms, via  $\psi
=\exp [-iz(\tilde{\alpha}_{x}+\tilde{\alpha}_{y}+k_0)/2]E $, and we assume $%
\chi \ll \tilde{\alpha}_{x},\tilde{\alpha}_{y}$ so we can perform
the so-called rotating wave approximation (see for instance
\cite{Moya}) to finally obtain
\begin{equation}
i\frac{\partial \psi }{\partial z}=\left( \tilde{\alpha}_{x}n_{x}+\tilde{%
\alpha}_{y}n_{y}+\tilde{\chi} (a_{x}a_{y}^{\dagger
}+a_{x}^{\dagger }a_{y})\right) \psi  , \label{two-fields2}
\end{equation}%
with
$\tilde{\chi}=\frac{\chi}{2k_0\sqrt{\tilde{\alpha_x}\tilde{\alpha_y}
}}$. This equation is equivalent to the field-field interaction in
quantum optics \cite{Dutra}.

\begin{figure}
\begin{center}
\includegraphics[width=0.4\textwidth]{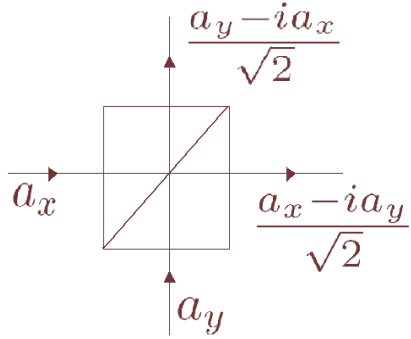}
\end{center}
\caption{\label{fig2}  Configuration of the beam splitter (50:50)
operation. }
\end{figure}

We do a last transformation $\phi =\exp
[-iz\tilde{\alpha}_{x}(n_x+n_y)]\psi $ and obtain
\begin{equation}
i\frac{\partial \phi }{\partial z}=[\Delta n_{y}+\tilde{\chi}
(a_{x}a_{y}^{\dagger }+a_{x}^{\dagger }a_{y})] \phi \equiv H\phi ,
\label{two-fields3}
\end{equation}%
with $\Delta=\tilde{\alpha}_y-\tilde{\alpha}_x$. It is useful to
define "normal-mode" operators by \cite{Dutra}
\begin{equation}
{A_1}=\delta a_x + \gamma a_y, \qquad A_2=\gamma a_x - \delta a_y,
\label{transformation}
\end{equation}%
with
\begin{equation}
\delta= \frac{2\tilde{\chi}}{\sqrt{2\Omega(\Omega-\Delta)}},
\qquad \gamma=\sqrt{\frac{\Omega-\Delta}{2\Omega}},
\qquad\delta^2+\gamma^2=1
\end{equation}%
with $\Omega=\sqrt{\Delta^2+4\tilde{\chi}^2}$ the Rabi frequency.
${A_1}$ and $A_2$ are annihilation operators just like $a$ and $b$
and obey the commutation relations
\begin{equation}
[{A_1},{A_1}^{\dagger}]=[A_2,A_2^{\dagger}]=1,
\end{equation}%
moreover, the normal-mode operators commute with each other
\begin{equation}
[{A_1},A_2]=[{A_1},A_2^{\dagger}]=0.
\end{equation}
In terms of these operator $H$ in  (\ref{two-fields3}) becomes
\begin{equation}
H=\mu_{1} A_1^{\dagger}{A_1} +\mu_{2} {A}_2^{\dagger}{A}_2,
\label{diag}
\end{equation} with $\mu_{1,2}= (\Delta \pm \Omega)/2$.
 In order
to have a way of transforming functions from one basis to the
other, we note that the lowest functions $u_0(x)u_0(y)$ are also
eigenfunctions of the normal-mode operators
\begin{equation}
A_mu_0(x)u_0(y)=0, \qquad m=1,2,
\end{equation}
i.e. they are the lowest states, up to a phase, in the new basis
\cite{Dutra}
\begin{equation}
u_0(x)u_0(y)= U_0^1(x,y)U_0^2(x,y).
\end{equation}
\subsubsection{Initial function $u_1(x)u_1(y)$}
If we consider
\begin{equation}
\phi(z=0;x,y)=u_1(x)u_1(y)
\end{equation}
the propagated function reads
\begin{equation}
\phi(z;x,y)=e^{-iz(\mu_{1} A_1^{\dagger}{A_1} +\mu_{2}
{A}_2^{\dagger}{A}_2)}a^{\dagger}_xa^{\dagger}_yu_0(x)u_0(y),
\end{equation}
by writing the creation operators in terms of creation operators
for the normal-modes, (\ref{transformation}), we obtain
\begin{equation}
\phi(z;x,y)=e^{-iz\mu_{1} A_1^{\dagger}{A_1}}e^{-iz \mu_{2}
{A}_2^{\dagger}{A}_2}(\delta A^{\dagger}_1+ \gamma
A^{\dagger}_2)(\gamma A^{\dagger}_1 -\delta
A^{\dagger}_2)U_0^1(x,y)U_0^2(x,y).
\end{equation}
Now we use the properties $e^{-iz\mu_{j} A_j^{\dagger}{A_j}}
A^{\dagger}_j e^{iz\mu_{j}
A_j^{\dagger}{A_j}}=A^{\dagger}_je^{-iz\mu_{j}}$ and
$e^{-iz\mu_{j} A_j^{\dagger}{A_j}}U^j_0(x,y)=U^j_0(x,y)$ to obtain
\begin{equation}
\phi(z;x,y)=(\delta A^{\dagger}_1 e^{-iz\mu_{1}}+ \gamma
A^{\dagger}_2e^{-iz\mu_{2}})(\gamma A^{\dagger}_1e^{-iz\mu_{1}}
-\delta A^{\dagger}_2e^{-iz\mu_{2}})U_0^1(x,y)U_0^2(x,y),
\end{equation}
that in the old basis reads
\begin{equation}
\phi(z;x,y)= [\eta(z)\beta(z) a^{\dagger
2}_x+\epsilon(z)\beta(z)a^{\dagger 2}_y+
(\beta^2(z)+\epsilon(z)\eta(z))a^{\dagger}_xa^{\dagger}_y]u_0(x)u_0(y),
\end{equation}
with $\eta(z)=\delta^2e^{-iz\mu_{1}}+\gamma^2e^{-iz\mu_{2}}$,
$\beta(z)=\gamma\delta (e^{-iz\mu_{1}}-e^{-iz\mu_{2}})$ and
$\epsilon(z)=\delta^2e^{-iz\mu_{2}}+\gamma^2e^{-iz\mu_{1}}$. The
term that multiplies $a^{\dagger}_xa^{\dagger}_y$ may be written
as
\begin{equation}
\nonumber e^{-i\Delta
z}[(\gamma^2-\delta^2)^2+4\gamma^2\delta^2\cos(\Omega z)],
\end{equation}
oscillates crossing zero periodically. By looking at the
propagated field at one of these zeros (at $z_0$), a state of the
form (\ref{2photons}) is obtained, i.e.
\begin{equation}
\phi(z_0;x,y)= \eta(z_0)\beta(z_0)u_2(x)u_0(y)+
\epsilon(z_0)\beta(z_0)u_0(x)u_2(y).
\end{equation}

\subsubsection{SU(2) coherent state}
One can engineer functions in the form of determined
superpositions of the functions (\ref{descomp}). For instance, one
can engineer the so-called SU(2) coherent state by setting the
functions $u_m(x)$ and $u_0(y)$ at the plane $z=0$
\begin{equation}
\phi(z=0;x,y)=u_m(x)u_0(y)= \frac{a_x^{\dagger m}}{\sqrt{m!}}
u_0(x)u_0(y),
\end{equation}
that after application of the propagator $e^{-iz(\mu_{1}
A_1^{\dagger}{A_1} +\mu_{2} {A}_2^{\dagger}{A}_2)}$ gives
\begin{equation}
\phi(z;x,y)=\frac{(\eta(z)a_x^{\dagger}+\beta(z)a_y^{\dagger})^m}{\sqrt{m!}}
u_0(x)u_0(y),
\end{equation}
that may be rewritten as the SU(2) coherent state \cite{Buzek,Wod}
\begin{equation}
\phi(z;x,y)=\sum_{n=0}^m \left(
\begin{array} {l}
m\\
n
\end{array} \right)^{1/2}
\eta^{m-n}(z)^n\beta(z) u_{m-n}(x)u_n(y).
\end{equation}
\subsubsection{Gaussian function}
Consider now that at $z=0$ we have a displaced Gaussian function
as a function of $x$ and a function $u_0(y)$,
\begin{equation}
\phi(z=0;x,y)=\left(\frac{k_0\tilde{\alpha}_x}{{\pi}}\right)^{1/4}{e^{-{(\alpha-x\sqrt{\frac{k_0\tilde{\alpha}_x}{2}})^2}}}u_0(y)\equiv
e^{-\frac{\alpha^2}{2}}\sum_n^{\infty}\frac{\alpha^n}{\sqrt{n!}}u_n(x)u_0(y),
\end{equation}
by using the Glauber displacement operator\cite{Glauber}, this
function may be rewritten as
\begin{equation}
\phi(z=0;x,y)=D_{a_x}(\alpha)u_0(x)u_0(y),
\end{equation}
with $D_{a_x}(\alpha)=\exp[\alpha(a_x^{\dagger}-a_x)]$. After
application of the propagator operator, the propagated field reads
\begin{equation}
\phi(z;x,y)=D_{a_x}[\alpha\eta(z)]u_0(x)
D_{a_y}[\alpha\beta(z)]u_0(y).
\end{equation}
The electromagnetic field then will be displaced Gaussians in $x$
and $y$ dimensions,  the initial Gaussians at $z=0$ will exchange
displacement, i.e. the displaced Gaussian (that depends on $x$)
will periodically diminish its displacement and back, while the
one centered at $y=0$  will periodically be displaced and back to
the origin.
\subsection{Conclusions}
We have shown how to model a quantum beam splitter by propagating
electromagnetic fields. Specific electromagnetic fields may be
engineered by using this modelling, in particular we have shown
how SU(2) coherent functions may be realized.

\end{document}